\documentclass[12pt,a4paper]{article}

\usepackage{jheppub}
\title{Large $N$ phase transitions in supersymmetric Chern-Simons theory  with massive matter}
\author[a]{Alejandro Barranco and}
\author[a,b]{Jorge G. Russo}
\affiliation[a]{ECM Department and Institute for Sciences of the Cosmos, Facultat de F\'\i sica, \\
Universitat de Barcelona, 
Mart\'i Franqu\`es 1, E08028 Barcelona, Spain.}
\affiliation[b]{Instituci\'o Catalana de Recerca i Estudis Avan\c cats (ICREA),\\
Pg. Lluis Companys 23, 08010 Barcelona, Spain.}
\emailAdd{alejandro@ecm.ub.edu}
\emailAdd{jorge.russo@icrea.cat}

\abstract{
We study   three-dimensional ${\cal N}=2$ $U(N)$ Chern-Simons theory on  $\mathbb S^3$ coupled to $2N_f$  chiral  multiplets   deformed by mass terms.
The partition function localizes to a matrix integral, which can be exactly computed in the large $N$ limit.
In a specific decompactification limit, the theory exhibits quantum (third-order) phase transitions at finite critical values of the  coupling. The theory presents  three  phases when $0<N_f<N$ and  two phases when $N_f\geq N$. The vacuum expectation value
of the supersymmetric circular Wilson loop has a discontinuity in the first derivative.
 }
\usepackage{color}

\begin{document}

\makebox[0pt][l]{\hspace*{120mm} \parbox{3cm}{ICCUB-14-002}}
\hspace*{-16mm}
\maketitle

%%%%%%%%%%%%%%%%%%%%%%%%%%%%%%%%%%%%%%%%%%%%%%%%%%%%%%%%%%%%%%%%%%%%%

\def\Xint#1{\mathchoice
   {\XXint\displaystyle\textstyle{#1}}%
   {\XXint\textstyle\scriptstyle{#1}}%
   {\XXint\scriptstyle\scriptscriptstyle{#1}}%
   {\XXint\scriptscriptstyle\scriptscriptstyle{#1}}%
   \!\int}
\def\XXint#1#2#3{{\setbox0=\hbox{$#1{#2#3}{\int}$}
     \vcenter{\hbox{$#2#3$}}\kern-.52\wd0}}
\def\ddashint{\Xint=}
\def\dashint{\Xint-}

\newcommand{\be}{\begin{equation}}\newcommand{\ee}{\end{equation}}
\newcommand{\bea}{\begin{eqnarray}} \newcommand{\eea}{\end{eqnarray}}
\def\sech{ {\rm sech}}
\def\p{\partial}
\def\pa{\partial}
\def\ov{\over }
\def\a{\alpha }
\def\g{\gamma}
\def\s{\sigma }
\def\td{\tilde }
\def\vp{\varphi}
\def\gd{\nu }
\def \ha {{1 \over 2}}

\def\KK{{\cal K}}

\newcommand\cev[1]{\overleftarrow{#1}}

%%\renewcommand{\thefootnote}{\fnsymbol{footnote}}
%%\setcounter{footnote}{0}
%%%%%%%%%%%%%%%%%%%%%%%%%%%%%%%%%%%%%%%%%%%%%%%%%%%%%%%%%%%%%%%%%%%%%

%\clearpage

%\setcounter{page}{1}
%\renewcommand{\thefootnote}{\arabic{footnote}}
%\setcounter{footnote}{0}

%\title{Large $N$ phase transitions in supersymmetric Chern-Simons theory with massive matter}
%\maketitle
%\tableofcontents
%%%%%%%%%%%%%%%%%%%%%%%%%%%%%%%%%%%%%%%%%%%%%%%%%%%%%%%%%%%%%%%%%%%%%

%%%%%%%%%%%%%%%%%%%%%%
\section{Introduction}
%%%%%%%%%%%%%%%%%%%%%%

Supersymmetric localization has led to the exact computation of the Euclidean partition function and vacuum expectation values  of Wilson loop operators in many supersymmetric gauge theories in various  dimensions. 
In the pioneering work by Pestun \cite{Pestun:2007rz}, the  method of localization was used to obtain exact formulas for  ${\cal N}=2$  super Yang Mills (SYM) theories on a four-sphere with arbitrary gauge group and matter content.
Soon after the method was  applied to the calculation of Euclidean path integrals in three-dimensional supersymmetric Chern-Simons theories on a three-sphere \cite{Kapustin:2009kz} and, since then, many other interesting examples have been worked out  (see \cite{Marino:2011nm} for a review and references and \cite{Beasley:2005vf} for earlier works).

For observables with a sufficient amount of supersymmetry, the final expressions are given in terms of a matrix integral. 
This integral is in general complicated, though   much simpler and much more under control
than the original functional integral. In the multicolor  limit, the integral is dominated by a saddle-point and in some cases the saddle-point equations can be solved exactly. 
Using this idea, the large $N$ behavior of the free energy and Wilson loops in ABJM theory were determined \cite{Drukker:2010nc},  leading to striking tests of the AdS/CFT correspondence. 
The large-$N$ master field of several four-dimensional ${\cal N}=2$ $U(N)$ super Yang Mills (SYM) theories has  also been determined (for a recent review and references, see \cite{Russo:2013sba}).
Among the different results that arise from this study, perhaps the most intriguing one  is the emergence of large $N$ quantum phase transitions \cite{Russo:2013qaa,Russo:2013kea},  which seem to be
generic features of  massive  ${\cal N}=2$ theories in the decompactification limit. This phenomenon was shown explicitly for  
${\cal N}=2^*$ SYM --obtained by  the unique mass deformation of ${\cal N}=4$ SYM preserving two supersymmetries-- and  ${\cal N}=2$ SQCD with $2N_f$  flavors, with $N_f<N$. 
Large $N$ phase transitions are familiar  in gauge theories and they are due to  singularities associated with the
finite radius of convergence of planar perturbation theory \cite{Gross:1980he,Wadia:2012fr}. However, for the supersymmetric observables computed in \cite{Russo:2013qaa,Russo:2013kea}, the physical origin of the phase transition appears to be different.  When the coupling crosses a  critical value, 
 field configurations with extra massless multiplets contribute to the saddle-point, leading to discontinuities in vacuum expectation values of supersymmetric  observables.

Similarly, one may expect that  massive   three-dimensional  ${\cal N}=2$ supersymmetric gauge theories on $\mathbb S^3$  also exhibit interesting large $N$ physics.
In particular,  one would like to know if  Chern-Simons theories coupled to massive matter  undergo quantum weak/strong coupling phase transitions. These questions can be again addressed by
using the exact results provided by localization \cite{Kapustin:2009kz} and matrix model techniques \cite{Drukker:2010nc, Marino:2011nm,Marino:2004eq}  (other
 studies of
Chern-Simons matter theories at large $N$ can be found in  e.g.  \cite{Herzog:2010hf,Santamaria:2010dm,Jafferis:2011zi}).

In this paper we study  the large $N$ limit of $U(N)$ Chern-Simons theory with $2N_f$ massive flavors. Like in the analogous four-dimensional case, we will find phase transitions arising in a specific decompactification limit of the theory.

%%%%%%%%%%%%%%%%%%%

%%%%%%%%%%%%%%%%%%%%%%%%%%%%%%%%%%%%%%%%%%%%%%%%%%%%%%%%%%
\section{$U(N)$ Chern-Simons with $2N_f$ massive flavors}
%%%%%%%%%%%%%%%%%%%%%%%%%%%%%%%%%%%%%%%%%%%%%%%%%%%%%%%%%%

Let us consider the ${\cal N}=2$ supersymmetric Chern-Simons theory with gauge group $U(N)$ on $\mathbb S^3$ and level $k$, with a matter content  given by
$2N_f$ chiral multiplets of mass $m$ ($N_f$ fundamentals and $N_f$ antifundamentals). For $m=0$, the theory is superconformal for any $N_f$ \cite{Schwarz:2004yj,Gaiotto:2007qi}. The mass  deformation for the chiral multiplets explicitly breaks classical scale invariance and hence conformal invariance.
Applying localization techniques,  one  finds that the exact functional integral of the partition function localizes to a finite dimensional integral over a subset of 
field configurations obeying classical equations and containing a one-loop determinant \cite{Kapustin:2009kz,Kapustin:2010xq}. In the conventions of \cite{Marino:2011nm}, the partition function localizes to the following matrix model integral \cite{Kapustin:2009kz,Kapustin:2010xq},  
\begin{equation}\label{Z}
Z^{U(N)}_{N_f} =
%\frac{i^{-\frac{N^2}{2}}}{N!} 
\int \frac{d^N \mu}{(2\pi)^N} \frac{\prod _{i<j} 4\sinh^2 (\frac{1}{2}(\mu_i-\mu_j))  
\ e^{-\frac{1}{2g} \sum_i  \mu_i^2}}{\prod_{i}\left(4 \cosh(\frac{1}{2}(\mu_i+m)) \cosh(\frac{1}{2}(\mu_i-m))\right)^{N_f}}
\ ,
\end{equation}
where  
\be
g=\frac{2\pi i}{k}\  .
\ee
$\mu_i/2\pi$ represent the eigenvalues
of the scalar field, $\sigma$, that belongs to  the three dimensional $\mathcal N=2$ vector multiplet and comes  from 
dimensional reduction of the gauge  field in the four dimensional $\mathcal N=1$ vector 
multiplet. The scalar field $\sigma$ has mass dimensions, 
therefore, in \eqref{Z} both $\mu$ and $m$ scale  with the radius of the three-sphere, $R$. 
The radius has  been set to one for notational convenience. The dependence on the radius will be restored
when considering the decompactification limit. 
Calculations will be performed for a real parameter $g>0$, which ensures the convergence of the integral. The dependence on $k$ can  be
recovered in the final expressions for the supersymmetric observables by analytic continuation.

In the infinite $N$ limit, the partition function can be determined  by a saddle-point calculation.
Here we will consider the Veneziano limit, where the 't Hooft coupling, 
\be
t\equiv gN\ ,
\ee 
and the Veneziano parameter, 
\be
\zeta\equiv \frac{N_F}{N} \ ,
\ee
 are kept fixed as $N\to \infty$.
It is useful to define the potential as
\begin{equation}\label{potential}
V(\mu_i)=\sum_{i=1}^N\left(\frac{\mu_i^2}{2} + gN_f\log\left[2 \cosh\frac{\mu_i+m}{2}\right] + gN_f\log\left[2 \cosh\frac{\mu_i-m}{2}\right]\right)\ .
\end{equation}
The saddle-point equations are then
\begin{equation}\label{der}
  \frac{1}{N}\sum_{j\neq i} \coth \frac{\mu_i-\mu_j}{2}=\frac{1}{t}V'(\mu_i)=
  \frac{\mu_i}{t}  +\frac{\zeta}{2} \tanh\frac{\mu_i +m}{2} + \frac{\zeta}{2}\tanh\frac{\mu_i -m}{2}
  \ .
\end{equation}

Introducing as usual the eigenvalue density
\begin{equation}
\label{rrr}
\rho(\mu)=\frac{1}{N}\sum_{i=1}^N\delta(\mu-\mu_i)\ ,
\end{equation}
 the saddle-point equation \eqref{der} is converted into a singular
integral equation:
\begin{equation}\label{saddleeq}
\dashint  d\nu\,\rho(\nu) \coth \frac{\mu-\nu}{2}=
  \frac{\mu}{t}  +\frac{\zeta}{2} \tanh\frac{\mu +m}{2} + \frac{\zeta}{2}\tanh\frac{\mu -m}{2}
  \ ,
\end{equation}
where the integral is defined by the principal value prescription.
This matrix model can be solved exactly. The solution is explicitly constructed in section \ref{sec:LargeN}. 
For clarity, we will first discuss the solution directly in the decompactification limit, where, as we will see, the model
exhibits the presence of quantum phase transitions. 

Another observable that can be  computed by localization is the vacuum expectation value (vev)
of the 1/2 supersymmetric circular Wilson loop \cite{Kapustin:2009kz,Gaiotto:2007qi},
\begin{equation}\label{Wilson}
W(C)=\left\langle\frac{1}{N}{\rm Tr}\ \mathcal P\exp\left(\oint_Cd\tau\,(iA_\mu \dot x^\mu+\sigma\vert\dot x\vert)\right)\right\rangle\ ,
\end{equation}
where the contour $C$ is the  big circle of $\mathbb S^3$.
%$A_\mu$ and $\sigma$ are the gauge and the scalar field coming from dimensional reduction of the 
%$\mathcal N=1$ vector multiplet in four dimensions.
The vev of the Wilson loop localizes to a matrix integral obtained
by replacing the fields by their classical values $A_\mu=0$ and $\sigma=\frac{1}{2\pi}\rm{diag}(\mu_1,\ldots,\mu_N)$,
\begin{equation}
W(C)=\left\langle \frac{1}{N}\sum_i e^{\mu_i}\right\rangle \ .
\end{equation}
In the large $N$ limit, this vacuum expectation value is just given by the average computed with the density function \eqref{rrr}, 
\begin{equation}
W(C)=\int d\mu \ \rho(\mu) \ e^{\mu} \ .
\end{equation}

%%%%%%%%%%%%%%%%%%%%%%%%%%%%%%%%%%%%%%%%%%%%%%%%%%%%%%%%%%%%%%%%%%%%%%%%%%%%%%%%%%%%%%%%%%%%%%%%%%%
\section{Large $N$ solution in the  decompactification limit}\label{sec:DecompactificationLimit}
%%%%%%%%%%%%%%%%%%%%%%%%%%%%%%%%%%%%%%%%%%%%%%%%%%%%%%%%%%%%%%%%%%%%%%%%%%%%%%%%%%%%%%%%%%%%%%%%%%%

Consider the integral equation  \eqref{saddleeq}.
The term $ \coth (\frac{1}{2}(\mu-\nu))$ represents a repulsive force among eigenvalues.
For $t>0$, the term $\mu/t$ is an harmonic force pushing the eigenvalues towards the origin.
The last two terms, proportional to $\tanh(\frac{1}{2}(\mu\pm m))$, are  forces pushing the 
eigenvalues towards $\mp m$, respectively.

If $t\gg 1$, the harmonic force is negligible. If, in addition, $m\gg 1$, then 
the potential is flat until $\mu =\mathcal O(m)$. As a result, the eigenvalues scale with $m$.
Restoring the dependence on the radius $R$ of $\mathbb S^3$, we can make this limit precise introducing  
the coupling $\lambda\equiv t/mR$ and taking the decompactification limit at fixed $\lambda$, i.e. 
\bea
&&m\rightarrow  mR\ , 
\qquad
\mu\rightarrow  \mu R\ ,
\qquad\text{with}\qquad R\rightarrow \infty
\nonumber
\\
&& t\equiv g N\to \infty\ ,\qquad \lambda\equiv\frac{ t}{mR} ={\rm fixed}\ .
\label{decompactification}
\eea
It is worth stressing that $t$ is dimensionless and a priori there is no reason why it should be scaled with
$mR$. However, if the decompactification limit is taken at fixed $t\ll mR$, then its only effect is to decouple the matter fields, as this is equivalent to sending the masses to infinity. 
This may be compared with  four-dimensional ${\cal N}=2$ SYM theory coupled to massive matter,
e.g. ${\cal N}=2^*$ SYM or ${\cal N}=2$ SCFT$^*$ which can be viewed as a UV regularization of pure
${\cal N}=2$ SYM theory \cite{Russo:2013kea}.
In that case, the limit of masses $M\to\infty $  at fixed 't Hooft coupling $\lambda $ does not decouple the
massive fields. In order to decouple the massive fields one needs to take at the same time $\lambda \to 0$
with fixed $MR\, e^{1\over \beta_0 \lambda}$, where $\beta_0<0$ is the one-loop $\beta $ function coefficient in $\beta _\lambda = \beta_0\lambda^2$. In other words, $\lambda\to 0 $ is required to renormalize
a one-loop divergence, viewing $M$  as  UV cutoff.
In Chern-Simons-matter theory, the 't Hooft coupling does not renormalize because it is proportional to a rational number, $ N/k$. Thus, in the limit $mR\to \infty $ with fixed $t$, matter fields are decoupled and the theory just flows to pure ${\cal N}=2$ Chern-Simons theory.
In what follows we will refer to ``decompactification limit" to the specific limit \eqref{decompactification}
where the most interesting physics arises. We will shortly see that this limit defines a regular limit of the theory.

We shall assume a one-cut solution where $\rho(\mu)$ is  supported in an interval $\mu\in[-A,A]$, with unit normalization,
\begin{equation}
\label{norr}
\int_{-A}^A\rho(\mu) d\mu =1\ .
\end{equation}
In the limit \eqref{decompactification}, the large $N$ saddle-point equation simplifies to
\begin{equation}\label{contsaddlelimit}
\int_{-A}^A d\nu\,\rho(\nu){\rm sign}(\mu-\nu)=\frac{\mu}{m\lambda}+\frac{\zeta}{2}\left({\rm sign}(\mu+m)+{\rm sign}(\mu-m)\right)\ ,
\end{equation}
where the dependence on $R$ has completely canceled out and  $\mu$, $m$ and $\lambda$ can now take arbitrary values. 

The solutions to  \eqref{contsaddlelimit} are different according to  the  value of the
coupling $ \lambda $. 
Consider first the case $0<\zeta < 1$. This gives rise to three phases.

\subsubsection*{Phase I: $\lambda < 1$}

This phase arises when  $A<m$, implying that
 $\vert\mu\vert< m$ for any $\mu$. Under these conditions, 
the sign functions on the right hand side of equation \eqref{contsaddlelimit} cancel  out. Flavors do not play any role and we find a uniform eigenvalue density:
\begin{equation}\label{rhosub}
\rho_{\rm I}(\mu)=\frac{1}{2m\lambda}\ ,
\end{equation}
supported in the interval $\mu\in[-m\lambda,m\lambda]$.

\subsubsection*{Phase II: $1<\lambda < (1-\zeta)^{-1}$ }

In this interval  of the coupling the eigenvalue density takes the form
\begin{equation}\label{rhoint}
\rho_{\rm II}(\mu)=\frac{1}{2m\lambda}+\frac{1}{2\lambda}\left(\lambda-1\right)\left(\delta(\mu+m)+\delta(\mu-m)\right)\ ,\qquad \mu\in[-m,m]\ ,
\end{equation}
with $A=m$. The coefficients of the  Dirac-$\delta $ functions are implied by the normalization condition \eqref{norr}, once $A=m$ is assumed.
A further justification of this solution requires a  regularization, which is provided automatically by the finite $R$ exact solution presented below.
 We shall return to this solution  in section \ref{sec:LargeN}.

\subsubsection*{Phase III: $\lambda > (1-\zeta)^{-1}$}

In this case the saddle-point equation is solved by the eigenvalue density
\begin{equation}\label{rhosup}
\rho_{\rm III}(\mu)=\frac{1}{2m\lambda}+\frac{\zeta}{2}\left(\delta(\mu+m)+\delta(\mu-m)\right)\ ,\qquad \mu\in[-m\lambda(1-\zeta),m\lambda(1-\zeta)]\ .
\end{equation}
This is the solution that one would obtain by formal differentiation of \eqref{contsaddlelimit} with respect to $\mu$.
In order for the $\delta$ functions to contribute to the integral in \eqref{contsaddlelimit},
we  must require $A>m$, i.e. $\lambda>(1-\zeta)^{-1}$.

\bigskip

The above three solutions $\rho_{\rm I}$, $\rho_{\rm II}$ and $\rho_{\rm III}$  will be reproduced in the next section by taking the decompactification limit in the general solution.
They  apply in three different
intervals of the coupling $\lambda$ and represent three different phases of the theory.

Thus, the picture is as follows. When $\lambda<1$, the eigenvalues are uniformly distributed in the interval 
$[-m\lambda,m\lambda]$. The width of the eigenvalue distribution therefore increases with $\lambda$, until $\lambda=1$, where
the eigenvalue distribution is extended in the interval $[-m,m]$.
Beyond $\lambda=1$, there is still a uniform distribution in the interval $[-m,m]$,
now with fixed width and a density that decreases as $1/\lambda $. At the same time, 
 some eigenvalues begin to accumulate at $\mu=\pm m$.
The width of the distribution stays fixed until $\lambda$ overcomes  $(1-\zeta)^{-1}$. Beyond this point, eigenvalues are uniformly distributed in an interval $[-m\lambda(1-\zeta),m\lambda(1-\zeta)]$, 
which expands as $\lambda$ increases,
but now with a fixed number $N_f$ of eigenvalues  accumulated at $\pm m$.

\bigskip

In the case $\zeta \geq 1$, i.e. $N_f\geq N$, the third phase disappears. The system has two phases I and II, represented by the solutions \eqref{rhosub}, \eqref{rhoint}, where now phase II holds in the interval $\lambda \in (1,\infty )$.

%In appendix \ref{app:FiniteN} we discuss the solution at finite $N$.

%%%%%%%%%%%%%%%%%%%%%%%%%%%%%%%%%%%%%%%%%%%%%%%%%%
\subsection{Free energy and critical behavior}
%%%%%%%%%%%%%%%%%%%%%%%%%%%%%%%%%%%%%%%%%%%%%%%%%%

The order of the phase transition is  defined as usual by  the analytic properties of  the free energy:
\begin{equation}
F=-\frac{1}{N^2}\log Z\ .
\end{equation}
We first consider $0<\zeta<1 $ and compute its derivative  with respect to the coupling, which is related to the second moment of the eigenvalue density, \begin{equation}\label{freeenergy}
\partial_\lambda F=-\frac{R}{2m\lambda^2}\langle\mu^2\rangle=\left\lbrace
\begin{array}{ll}
-\frac{mR}{6}&\text{Phase I}
\\[1mm]
-\frac{mR}{6\lambda^3}(3\lambda-2 )&\text{Phase II}
\\[1mm]
-\frac{mR}{6\lambda^2}\left(\lambda^2(1-\zeta)^3+3\zeta\right)&\text{Phase III}  
\end{array}
\right.
\end{equation}
This implies a discontinuity in the third derivative at both critical points, $\lambda=1$ and $\lambda=(1-\zeta)^{-1}$:
\begin{equation}
\partial^3_\lambda(F_{\rm I}-F_{\rm II})\Big\vert_{\lambda=1}=-mR\ ,\qquad 
\partial^3_\lambda(F_{\rm II}-F_{\rm III})\Big\vert_{\lambda=(1-\zeta)^{-1}}=mR(1-\zeta)^5\ .
\end{equation}
Therefore, both phase transitions are third order. 
The  free energy in the three phases is  given by: 
\begin{align}
F_{\rm I} &=\frac{mR}{6}(6\zeta -\lambda)\ ,
\\ 
F_{\rm II}&=\frac{mR }{6\lambda^2}
\left(3(2 \zeta -1) \lambda^2+3 \lambda -1\right)\ ,
\\
F_{\rm III}&=\frac{mR}{6\lambda}
\left((\zeta-1)^3\lambda ^2+3 \zeta ^2 \lambda  +3\zeta \right)\ ,
\end{align}
up to a common numerical constant.  Note that the free energy is complex upon analytic continuation to imaginary $g$. This is expected as the partition function \eqref{Z} with imaginary $g$ is complex.

In the case $\zeta\geq 1$, the expressions for the free energies $F_{\rm I}$ and $F_{\rm II}$ are the same,
but, as explained, phase III disappears and phase II extends up to $\lambda=\infty $.

%%%%%%%%%%%%%%%%%%%%%%%%%%%
\subsection{Wilson loop}
%%%%%%%%%%%%%%%%%%%%%%%%%%%

We now compute \eqref{Wilson} in the large $R$ limit  using the density functions \eqref{rhosub}, \eqref{rhoint} and \eqref{rhosup}. 
We obtain ($0<\zeta <1$)
\begin{equation}\label{Wilsonloop}
W(C)=\langle e^{\mu R} \rangle \sim \left\lbrace
\begin{array}{ll}
e^{mR\lambda}&\text{Phase I}
\\[2mm]
e^{mR}&\text{Phase II}
\\[2mm]
e^{mR\lambda(1-\zeta)} &\text{Phase III}
\end{array}\right.
\end{equation}
It follows a perimeter law, just like in massive (or asymptotically free) four-dimensional ${\cal N}=2$ SYM theories  \cite{Russo:2013qaa,Russo:2013kea,Russo:2012ay,Buchel:2013id}. At the two critical points, 
\begin{equation}
\partial_\lambda \log W(C)\sim \left\lbrace
\begin{array}{ll}
mR&\text{Phase I}
\\[2mm]
0&\text{Phase II}
\\[2mm]
mR(1-\zeta) &\text{Phase III}
\end{array}\right.
\end{equation}
Thus there is a discontinuity in the first derivative.\footnote{Power-like factors in $W$ \eqref{Wilsonloop} are not meaningful, since they are affected by subleading corrections which were discarded in the saddle-point equation \eqref{contsaddlelimit}.
A formal calculation  using the densities \eqref{rhosub}-\eqref{rhosup} including the power factors gives a $W$  with discontinuities in the second derivatives. The discontinuity in the first derivative then appears in the infinite $R$ limit.}

%%%%%%%%%%%%%%%%%%%%%%%%%%%%%%%%%%%%%%%%%%%%%%%%%
\section{Large $N$ solution at finite $R$}\label{sec:LargeN}
%%%%%%%%%%%%%%%%%%%%%%%%%%%%%%%%%%%%%%%%%%%%%%%%%

\subsection{General solution}

The integral equation \eqref{saddleeq} can be solved in general for finite $R$
 using standard methods \cite{Marino:2011nm,Marino:2004eq}.  
%To use the results explained in \cite{Marino:2011nm,Marino:2004eq}, we should
%have the standard Vandermonde determinant inside the partition
%function \eqref{Z}, to this purpose, 
It is convenient to make the following change of integration variables:
\begin{equation}\label{changeofvariable}
z_i =ce^{\mu_i}\ ,\qquad c\equiv e^{t(1-\zeta)}\ .
\end{equation}
Now we use the relations:
\begin{align}
d^N \mu\prod_{i<j}  4\sinh^2 \frac{\mu_i-\mu_j}{2} &= d^N z 
\frac{\prod _{i<j} (z_i-z_j)^2}{ \prod_{i} z_i^N}\ ,
\\
\prod_i \left(4 \cosh\frac{\mu_i+m}{2} \cosh\frac{\mu_i-m}{2}\right) &= 
c^{N} \prod_i z_i^{-1} \left(1+z_i \frac{e^{+m}}{c}\right)\left(1+z_i \frac{e^{ -m}}{c}\right)\ ,
\end{align}
The partition function becomes
\begin{equation}
Z^{U(N)}_{N_f} = e^{-\frac{t}{2} N^2(1-\zeta^2)} \int d^N z \ \prod _{i<j} (z_i-z_j)^2  \ e^{-\frac{1}{g} \sum_i  V(z_i)}\ , 
\end{equation}
which now exhibits a factor representing the Vandermonde determinant. The potential is
 given by
\begin{equation}
V(z)=\frac{1}{2}(\log z)^2 +  t\zeta  \log\left[ \left(1+ z \frac{e^{+m}}{c}\right)\left(1+z \frac{e^{ -m}}{c}\right)\right]\ .
\end{equation}
Therefore, we have a usual matrix model with logarithmic terms in the potential.
In these new variables, the saddle-point equation becomes
\begin{equation}
\dashint_a^b dz\,\hat\rho(z) \frac{1}{p-z}= \frac{1}{2t}\, V'(p)\ ,
\end{equation}
where $\hat\rho(z)dz=\rho(\mu)d\mu$.
To compute the eigenvalue density one defines the auxiliary ``resolvent'' function as
\begin{equation}
\omega (p) =\frac{1}{N} \left\langle \sum_{i=1}^N \frac{1}{p-z_i} \right\rangle\ ,
\end{equation}
whose  expression in the large $N$ limit is 
\begin{equation}\label{resolvent}
\omega(p)=\int dz\frac{\hat \rho(z)}{p-z}\ .
\end{equation}
For a generic potential $V(z)$, the resolvent  is then given by 
\cite{Marino:2004eq,Marino:2011nm}
\begin{equation}\label{resolventsol}
\omega(p) = \frac{1}{2t} \oint_{\cal C} \frac{dz}{2\pi i} \frac{V'(z)}{p-z} \left( \frac{(p-a)(p-b)}{(z-a)(z-b)}\right)^{1/2}\ ,
\end{equation}
where $\mathcal C$ is a path enclosing the branch cut defined by the branch points $a$ and $b$.\footnote{Multi-cut solutions are not supported by the numerical results.}
Then  the eigenvalue density is obtained from  the discontinuity of the resolvent accross the branch cut,
\begin{equation}
\hat \rho(p)=-\frac{1}{2\pi i}\left(\omega(p+i\epsilon)-\omega(p-i\epsilon)\right)\ .
\end{equation}
Equation \eqref{resolventsol} leads to
\begin{equation}
\omega(p) =\frac{1}{2t} V'(p) -\frac{1}{2t} M(p) \sqrt{(p-a)(p-b)}\ ,
\end{equation}
with
\begin{equation}
M(p)= \oint_{\infty} \frac{dz}{2\pi i} \frac{V'(z)}{z-p} \frac{1}{\sqrt{(z-a)(z-b)}}\ ,
\end{equation}
where the integral is done over the same path $\mathcal C$, but enclosing the point at infinity.

The integral defining $M(p)$ contains two contributions, $M=M_1+M_2$: $M_1$ coming from the potential term $(\log z)^2$, which is the one that appears in the pure Chern-Simons matrix model.
This integral --computed in \cite{Marino:2004eq}-- gives
\begin{equation}
M_1(p) =\frac{1}{p \sqrt{(p-a) (p-b)}}\log \frac{\big(\sqrt{a} \sqrt{p-b}-\sqrt{b} \sqrt{p-a}\big)^2}{p \left(\sqrt{p-a}-\sqrt{p-b}\right)^2}+\frac{2}{p\sqrt{ab}}\log \frac{\sqrt{a}+\sqrt{b}}{2\sqrt{a b}}\ .
\end{equation}
The second piece $M_2$ is 
\begin{equation}
M_2(p)= t\zeta   \oint_{\infty} \frac{dz}{2\pi i} \frac{1}{z-p} \frac{1}{\sqrt{(z-a)(z-b)}} 
\left(\frac{1}{c e^m+ z } +\frac{1}{c e^{-m}+ z }\right)\ .
\end{equation}
There is no contribution from the residue at $z=\infty$,  the only contributions come from the simple poles
at $z= -c e^{\pm m}$.
We find
\begin{equation}
M_2(p) =- t\zeta \left(\frac{1}{p+ce^m} \frac{1}{\sqrt{(a+c e^m)(b+ce^m)}} + (m\leftrightarrow -m ) \right)\ .
\end{equation}
Let us combine this with the contribution coming from the $(\log z)^2$ term. We write
$\omega =\omega^{(1)}+\omega^{(2)}$, where
\begin{align}
&\omega^{(1)}(p)= 
-\frac{1}{2tp} \log \frac{\big(\sqrt{a}\sqrt{p-b} -\sqrt{b}\sqrt{p-a} \big)^2}{p^2\left(\sqrt{p-a} -\sqrt{p-b}\right)^2}
-\frac{\sqrt{(p-a)(p-b)}}{tp \sqrt{ab}} \log \frac{\sqrt{a}+\sqrt{b}}{2\sqrt{ab}}\
, \label{omega1}
\\
&\omega^{(2)}(p)= \frac{\zeta}{2}\left(\frac{1}{c e^m+ p } +\frac{1}{c e^{-m}+ p }\right)-\frac{1}{2t} M_2(p) \sqrt{(p-a)(p-b)}\ . \label{omega2}
\end{align}
According to \eqref{resolvent}, the resolvent obeys the following boundary condition:
\begin{equation}
\omega(p)\sim \frac{1}{p}\ ,\qquad\text{for}\qquad p\to \infty\ .
\ee
Imposing this asymptotic condition to the solution \eqref{omega1} and \eqref{omega2},
we obtain two equations that determine the branch points $a$ and $b$,
\begin{align}
0&=\frac{\zeta}{2} \left( \frac{1}{\sqrt{(a+c e^m)(b+ce^m)}} + (m\leftrightarrow -m ) \right)-\frac{1}{t \sqrt{ab}} \log\frac{\sqrt{a}+\sqrt{b}}{2\sqrt{ab}}\ , 
\label{p0}
\\
1&=\zeta-\frac{\zeta}{2}\left(\frac{ce^{m}+\frac{1}{2}(a+b)}{\sqrt{(a+c e^m)(b+ce^m)}}+ (m\leftrightarrow -m )\right)
\nonumber\\
&\phantom{=\ }+\frac{(\sqrt{a}+\sqrt{b})^2}{2t\sqrt{ab}}\log\frac{\sqrt{a}+\sqrt{b}}{2\sqrt{ab}}+\frac{1}{t}\log\sqrt{ab}\ .
\label{p1}
\end{align}
Now, using the reflection symmetry of the original potential \eqref{potential} prior to the change of variable \eqref{changeofvariable},
we find that $a$ and $b$ obey the relation,
\begin{equation}
ab=c^2\equiv e^{2t(1-\zeta)}\ .
\end{equation}
As a result, one of  the two equations \eqref{p0} or \eqref{p1} becomes redundant. 
The solution for the eigenvalue density  takes the form
\begin{align}\label{rho}
\hat \rho(z)&=\frac{1}{ \pi tz }\frac{\sqrt{z-a} \sqrt{b-z}}{\sqrt{ab}}\log\left(\frac{\sqrt{a}+\sqrt{b}}{2\sqrt{ab}}\right)+\frac{1}{\pi tz}\tan ^{-1}\left(\frac{\sqrt{z-a} \sqrt{b-z}}{z+\sqrt{ab}}\right)
\nonumber\\
&\phantom{=\ }-\frac{\zeta}{2\pi}  \left(\frac{\sqrt{z-a} \sqrt{b-z}}{(c e^m+z)\sqrt{a+ce^m} \sqrt{b+ce^m} }+(m\rightarrow -m)\right)\ ,
\end{align}
with $z\in(a,b)$, $b=c^2a^{-1}$ and $a$ defined by one of the conditions
\eqref{p0} or \eqref{p1}.

The expression for the eigenvalue density takes a simpler form in terms of the original $\mu$ variable:
\begin{equation}\label{rhomu}
\rho(\mu)=\frac{1}{\pi  t}\tan ^{-1}\left(\frac{\sqrt{\cosh A-\cosh\mu}}{\sqrt{2}\cosh\frac{\mu}{2}}\right)
+\frac{\zeta}{\pi}\frac{ \cosh \frac{\mu}{2}\cosh \frac{m}{2}}{\cosh \mu+\cosh m}
\frac{\sqrt{\cosh A -\cosh\mu}}{\sqrt{\cosh A +\cosh m}}
\end{equation}
supported on the interval $\mu\in(-A,A)$, where $A$ is given by
the condition
\begin{equation}\label{A}
\log \left(\cosh\frac{A}{2}\right)=\frac{1}{2}t(1-\zeta)+\frac{t\zeta\cosh\frac{m}{2}}{\sqrt{2}\sqrt{\cosh A+\cosh m}}\ ,
\end{equation}
for any $\zeta\geq 0$.

In the massless  $m=0$ case, the eigenvalue density becomes
\begin{equation}\label{rhom0}
\rho(\mu)=\frac{1}{\pi  t}\tan ^{-1}\left(\frac{\sqrt{\cosh A-\cosh\mu}}{\sqrt{2}\cosh\frac{\mu}{2}}\right)
+\frac{\zeta}{2\pi}
\sqrt{\sech^2 \frac{\mu}{2} -\sech^2 \frac{A}{2} }
\end{equation}
\begin{equation}\label{Am0}
\log X= -\frac{t}{2} (1-\zeta + \zeta X)\ ,\qquad X\equiv \sech \frac{A}{2} \ .
\end{equation}
In particular, if $\zeta = 0$, i.e. pure ${\cal N}=2$ CS theory without matter, this reproduces 
the result of \cite{Marino:2011nm,Marino:2004eq}. This provides a check of our assumption that, for real $g$,  eigenvalues lie on one  cut in the real axes. For imaginary $g$, the cut lies in the complex plane.\footnote{It is simpler to perform the  continuation to imaginary $g$ after computing observables.}

\bigskip

As the coupling $t$ is gradually increased from 0, the eigenvalue density behaves as follows. At weak coupling, the classical force term $\mu/t$ in the saddle-point equation \eqref{saddleeq} is  dominant, squeezing the eigenvalue distribution
towards the origin. All eigenvalues are small and  the kernel in the integral of equation \eqref{saddleeq} approaches the Hilbert kernel, leading to the  Wigner semicircular distribution,
\begin{equation}
\rho(\mu)\approx \frac{1}{2\pi t}\sqrt{4t-\mu^2}\qquad \mu\in\left[-2\sqrt{t},2\sqrt{t}\right]\
,\qquad t\ll 1\ .
\end{equation}
Indeed, this expression can be obtained directly  from \eqref{rhomu}. In fig. \ref{Wigner} we show this distribution as compared to the finite $N$ eigenvalue density obtained numerically from eq. \eqref{der}.
% (for finite $N$, we adopt the definition  \eqref{discreterho}).
\begin{figure}[h!]
\centering
\includegraphics[width=0.4\textwidth]{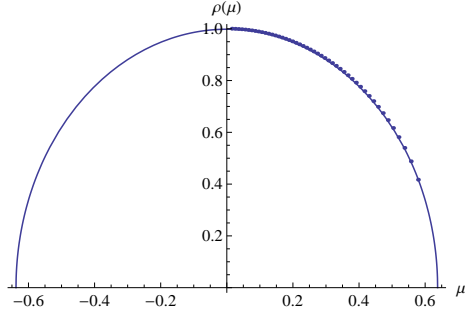}
\caption{At $t\ll 1 $ the eigenvalue density approaches the Wigner distribution ($t=0.1$, $m=50$, $\zeta=0.25$). Solid line: eigenvalue distribution obtained analytically. Dots: numerical solution to \eqref{der}  with $N=100$. 
}
\label{Wigner}
\end{figure}

As  $t$ is further increased, the eigenvalue distribution expands and gets flattened forming a plateau, until $t$ gets close to $t\lesssim m$, when two peaks around $\mu\approx\pm m$
begin to form (fig. \ref{subcr}). For finite $R$,  small peaks begin to show up already at $t\lesssim m$.

\begin{figure}[h!]
\centering
\begin{tabular}{cc}
\includegraphics[width=0.4\textwidth]{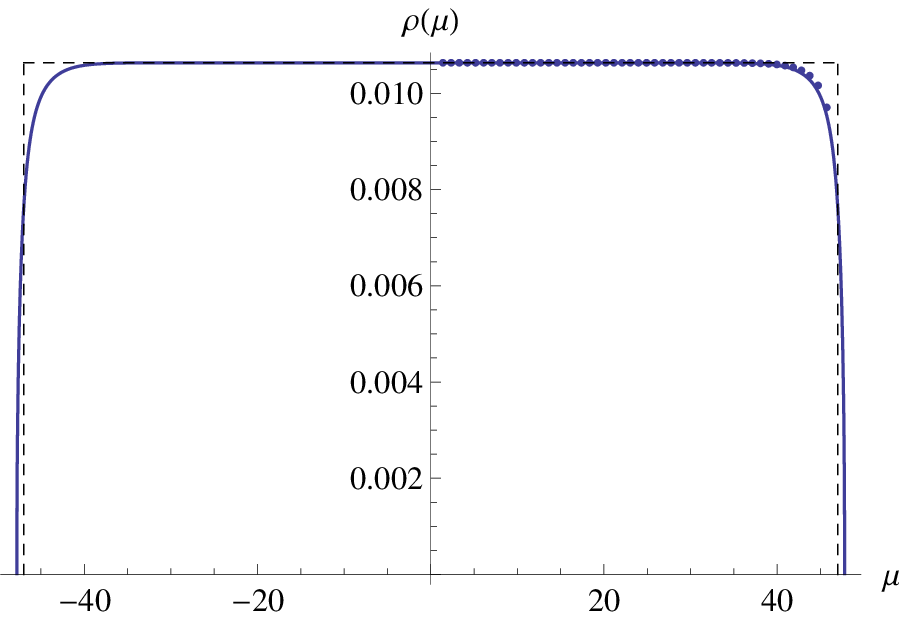}
&
\includegraphics[width=0.4\textwidth]{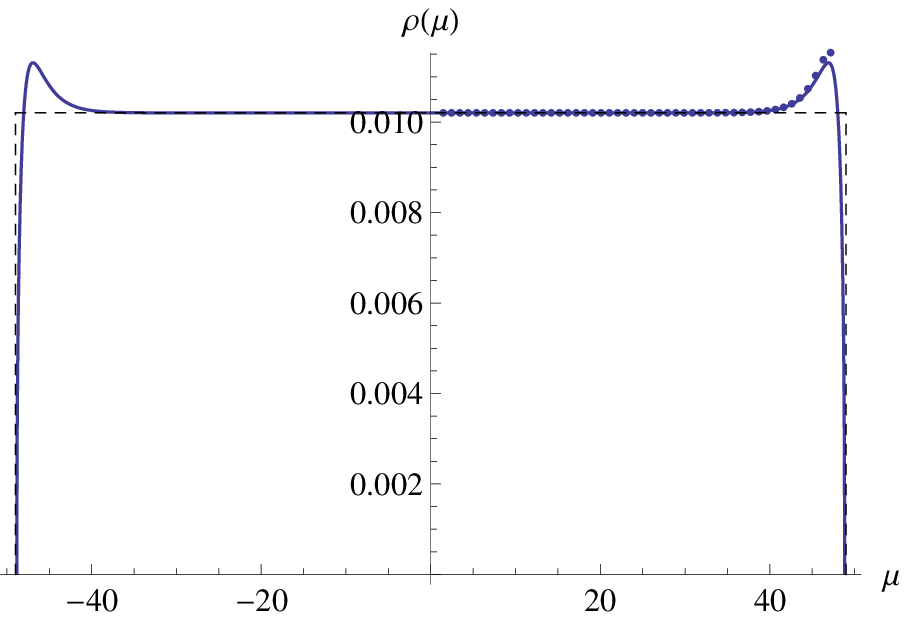}
\\
(a)&(b)
\end{tabular}
\caption{Eigenvalue density in phase I for $m=50$, $\zeta=0.25$ and (a) $t=47$, (b) $t=49$.
Solid line:  analytic solution. Dashed line: solution in the decompactification limit. Dots: numerical solution to \eqref{der}  with $N=100$.}
\label{subcr}
\end{figure}

As the coupling is  increased beyond $t=m$, 
 eigenvalues begin to accumulate around  $\mu=\pm m$, enhancing the peaks and maintaining the plateau between them (this is shown in fig. \ref{intermediate}). This
would correspond to phase II in the decompactification limit, where peaks
turn into Dirac delta functions. For $\zeta\geq 1$ this phase holds up to $t=\infty$:
the eigenvalue distribution  is uniform with support in a fixed interval $(-m,m)$, with a density decreasing as $1/t$,  and with two peaks at $\mu=\pm m$, whose amplitudes increase until all eigenvalues get  on the top of $\mu =\pm m$ as $t\to\infty$.

\begin{figure}[h!]
\centering
\begin{tabular}{ccc}
\includegraphics[width=0.4\textwidth]{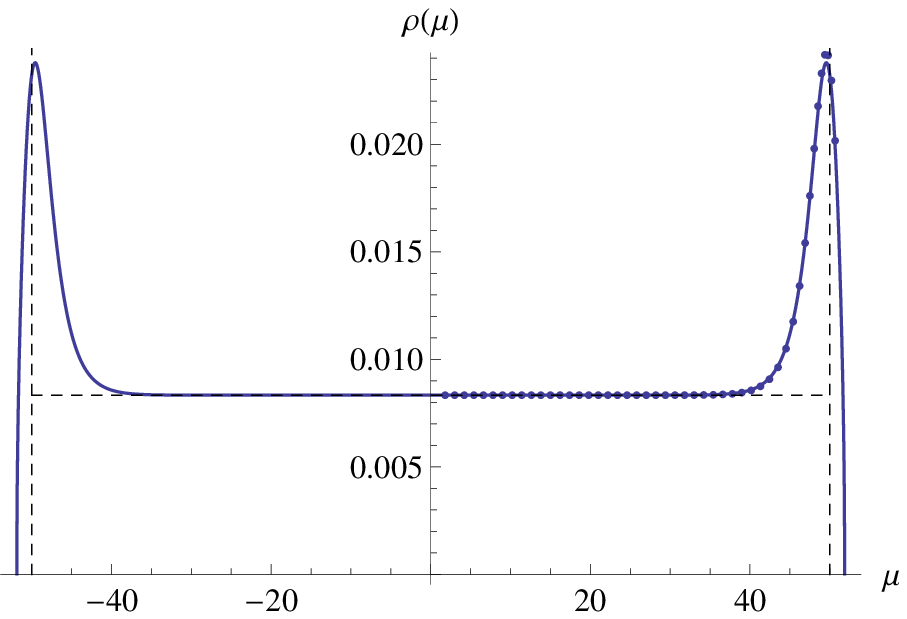}
&
\includegraphics[width=0.4\textwidth]{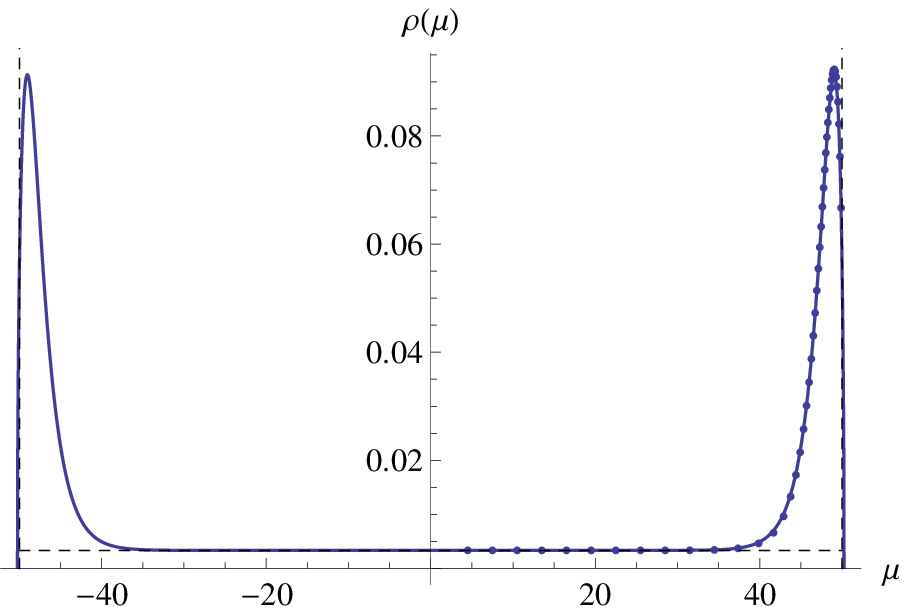}
\\
(a)&(b)
\end{tabular}
\caption{Eigenvalue density in phase II for $m=50$ and  (a) $t=60$, $\zeta=0.25$, (b) $t=150$, $\zeta=2$
(same conventions as in fig. \ref{subcr}).}
\label{intermediate}
\end{figure}

When $0<\zeta< 1$, phase II holds
only in the interval $m<t<m/(1-\zeta)$.
For $t>m/(1-\zeta)$, the plateau begins to extend beyond the peaks at $\mu=\pm m$, as shown
in fig. \ref{supcr}. 
Each peak now contains $N_f/2$ eigenvalues.
This  reproduces the behavior found   in section \ref{sec:DecompactificationLimit} for phase III.

Note that fig.\ref{intermediate}b and \ref{supcr} display the eigenvalue density for
the same value of $t=150$ but different $\zeta $. They illustrate the fact that
when $\zeta \geq 1$ eigenvalues lie on the interval $[-m,m]$ for all $t>m$,
whereas when $\zeta <1$ the eigenvalue distribution extends beyond $\mu=\pm  m$ 
as soon as $t$ overcomes $m/(1-\zeta)$.

\begin{figure}[h!]
\centering
\includegraphics[width=0.4\textwidth]{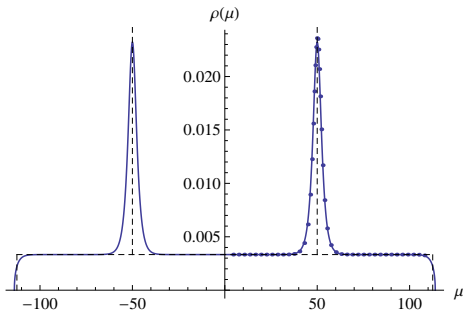}
\caption{Eigenvalue density in phase III for $m=50$, $t=150$, $\zeta=0.25$ (same conventions as in fig. \ref{subcr}).}
\label{supcr}
\end{figure}

\bigskip

Using the eigenvalue density \eqref{rhomu}, we can obtain the expression for the Wilson loop at finite $R$,
\begin{align}
W(C)&=\frac{1}{t}\sinh^2\frac{A}{2}+\frac{\zeta}{2}\frac{\sqrt{1+\cosh m}}{\sqrt{\cosh A+\cosh m}}
\nonumber\\
&\phantom{=\ }\times
\left(\cosh A-1+2\cosh m\left(1-\frac{\sqrt{\cosh A+\cosh m}}{\sqrt{1+\cosh m}}\right)\right)\ .
\end{align}

%%%%%%%%%%%%%%%%%%%%%%%%%%%%%%%%%%%%%%%%%%%%%%%%%%%%%%%%%%%%%%%
\subsection{Decompactification limit}
%%%%%%%%%%%%%%%%%%%%%%%%%%%%%%%%%%%%%%%%%%%%%%%%%%%%%%%%%%%%%%%

%Taking the decompactification limit in section \ref{sec:LargeN}
%we should recover the results of section \ref{sec:DecompactificationLimit}. 

Let us  examine the general formula for the eigenvalue density \eqref{rhomu}, \eqref{A} in the large $R$ limit.
It is convenient to restore the $R$ dependence by the scaling $m\to mR$, $A\to AR$, $\mu\to \mu R$. 
For large $R$, 
  \eqref{A} simplifies to the following form
\begin{equation}\label{Alimit}
A-\frac{1}{R}\log 4=m\lambda(1-\zeta)+\frac{m\lambda\zeta}{\sqrt{e^{(A-m)R}+1}}\ ,
\end{equation}
where, again,  we have introduced the parameter $\lambda\equiv t/mR$.
We now solve this equation in the three different phases:

\begin{itemize}

\item $\lambda<1$:  Let us assume that $A<m$. In this case we can neglect the exponential inside the square root of \eqref{Alimit}. This gives
 $A\approx m\lambda$. Thus the $A<m$ phase appears when $\lambda<1$.

\item $1<\lambda<(1-\zeta)^{-1}$: In this interval the solution is of the form:
\be
\label{asw}
A = m + \frac{1}{R}  \log\left[ \frac{ \lambda^2 \zeta^2}{  (1- \lambda(1-\zeta)) ^2} -1\right]+ 
\mathcal O(R^{-2})\ .
\ee
As we will shortly see, the $\mathcal O(R^{-1})$ term is important in determining the density at $R\to \infty$.
When $\zeta \geq 1$, this solution for $A$ is real for any $\lambda>1 $, and in this case this phase extends up to $\lambda =\infty$.
When $0<\zeta < 1$, \eqref{asw} solves \eqref{Alimit} with real $A$  provided $1<\lambda<(1-\zeta)^{-1}$.

\item $\lambda>(1-\zeta)^{-1}$: Let us now assume that $A>m$. In this case the last term of eq. \eqref{Alimit} can be neglected and we end up with
\begin{equation}
A\approx m\lambda(1-\zeta)\ .
\end{equation}
Thus the solution arises only when $\zeta<1$ and $A>m$ requires $\lambda>(1-\zeta)^{-1}$, in concordance  with the analysis of section \ref{sec:DecompactificationLimit}.

\end{itemize}

\noindent Consider now the eigenvalue density \eqref{rhomu}.
 The first term gives
\begin{equation}
\frac{1}{\pi  m\lambda}\tan ^{-1}\left(\frac{\sqrt{\cosh AR-\cosh\mu R}}{\sqrt{2}\cosh\frac{\mu R}{2}}\right)\mathop{\longrightarrow}_{R\rightarrow\infty} 
\left\lbrace\begin{array}{lc}
0& ,\ \ \vert\mu\vert = A\ ,
\\
\frac{1}{2m\lambda}& ,\ \ \vert\mu\vert < A\ .
\end{array}\right.
\end{equation}
Therefore this is the term which gives the plateau,  reproducing the same result of 
section \ref{sec:DecompactificationLimit}.

Consider now the second term in \eqref{rhomu}.
When $A<m$, this term vanishes at large $R$.
If, instead, $A>m$, then this term  generates two Dirac delta functions centered on $\pm m$
with normalization $\zeta/2$. 
For a trial function $f(\mu)$,  one numerically finds that
\begin{equation}
\label{afas}
R\int_{-A}^A d\mu\,\frac{2}{\pi}\frac{ \cosh \frac{\mu R}{2}\cosh \frac{mR}{2}}{\cosh \mu R+\cosh mR}
\frac{\sqrt{\cosh AR -\cosh\mu R}}{\sqrt{\cosh AR +\cosh mR}}\ f(\mu )\ \longrightarrow \ f(m )+f(-m)\ ,
\end{equation}
at large $R$.

Finally, consider the intermediate case, phase II, where $A$ is given by \eqref{asw}.  We find a similar result as \eqref{afas}, but with an extra overall coefficient $  (\lambda-1)/(\zeta \lambda)$. This coefficient is produced by the correction of order $\mathcal O(R^{-1})$ in $A$.
Thus the resulting $\rho$ exactly matches  the solution \eqref{rhoint}.

%%%%%%%%%%%%%%%%%%%%%
%\section{Conclusions}
%%%%%%%%%%%%%%%%%%%%%

%Mass deformed three-dimensional supersymmetric field theories exhibit mirror symmetry under the %exchange
% of mass and Fayet-Iliopolous (FI) parameters \cite{Kapustin:2010xq}. It would be interesting to study 
% the consequences of this interplay may shed a new understanding on the origin of the phase transitions.
% Finally, we comment on mass deformed ABJM theory.

\bigskip

In summary, like in ${\cal N}=2$ massive four-dimensional SYM theories, mass deformations in ${\cal N}=2$  supersymmetric three-dimensional Chern-Simons-matter theory 
lead to new physics involving large $N$ quantum phase transitions. These phase transitions produce non-analytic behavior in supersymmetric observables, like discontinuities in the first derivatives of the vev of the circular Wilson loop, which can be computed explicitly.
In this paper we have not included Fayet-Iliopoulos (FI) parameters.
Including both FI and mass parameters may shed  new light  on the properties of the phase transitions.
  The exchange
of mass and FI parameters exchanges mirror pairs of
three-dimensional supersymmetric field theories \cite{Kapustin:2010xq,Benvenuti:2011ga}. In particular, this indicates that certain massless theories deformed by FI parameters
may also exhibit large $N$ phase transitions in some limit.
It would be interesting to study the consequences of this interplay in more detail.
It would also be interesting to study the analogous decompactification limit in the mass-deformed ABJM 
partition function given in \cite{Kapustin:2010xq}.

%%%%%%%%%%%%%%%%%%%%%%%%%%%%%%
\subsection*{Acknowledgements}
%%%%%%%%%%%%%%%%%%%%%%%%%%%%%%
 
We are grateful to K. Zarembo and M. Mari\~no for  useful  comments. 
The numerical results shown in section \ref{sec:LargeN}
were obtained by adapting a Mathematica code developed by Zarembo to the present models.
We acknowledge financial support from projects FPA 2010-20807.
% and CPAN (Consolider CSD2007-00042).
A.B. also acknowledges support from MECD FPU fellowship
AP2009-3511. 
%%%%%%%%%%%%%%%%%%%%%%%%%%%%%%

\end{document}